# Energy-Efficient Cell Partition of 3D Space for Sensor Networks with Location Information

Susumu Matsumae

Dept. of Information Science, Saga University

Saga 840-8502, Japan

E-mail: matsumae@is.saga-u.ac.jp

**Abstract**

In the wireless sensor networks composed of battery-powered sensor nodes, one of the main issues is how to save power consumption at each node. The usual approach to this problem is to activate only necessary nodes (e.g., those nodes which compose a backbone network), and to put other nodes to sleep. One such algorithm using location information is GAF (Geographical Adaptive Fidelity). GAF is enhanced to HGAF (Hierarchical Geographical Adaptive Fidelity). In this paper, we study the energy-efficient partition of a 3 dimensional sensor field into cells. Further, we give a theoretical upper bound on the cell size for this problem.

**Keywords:** Wireless sensor network, Geographical adaptive fidelity, Energy conservation, Network lifetime, Location information,

## 1. Introduction

Wireless sensor networks have gained much attention in recent research and development. In the wireless sensor networks, battery-powered sensor nodes are placed on the observation area, and the sensed data is transmitted to the observer by multi-hop communication between nodes. Traditionally, the routing protocols for these networks have been evaluated in terms of packet loss rates, routing overhead, etc. However, since wireless sensor networks are usually deployed using battery-powered nodes, the optimization of routing protocol's energy consumption is also important [1][2][3].

The usual technique for designing an energy-efficient routing protocol is to activate only necessary nodes (e.g., those nodes which compose a backbone network) and to put other nodes to sleep [4][5][6][7][8][9][10][11]. Among these protocols, in this paper we focus on GAF (Geographical Adaptive Fidelity) [11] and its extended version called HGAF (Hierarchical Geographical Adaptive Fidelity) [6][7][8][9]. In the GAF-based algorithms, the sensor field is partitioned by regions called *cells*, and the cell size affects the energy efficiency of the protocols. Table 1 summarizes the maximum cell sizes for the GAF-based methods that use regular polygon as cell shapes. As shown in Table 1, we successfully obtained the cell size of $(3\sqrt{3}/4)R^2$ [7], which is 29.9% larger than that of eHGAF whose cell size is $R^2$ [6].

Table 1. Maximum cell sizes for GAF, HGAF, eHGAF, and eHGAF with triangle cells. ($R$ denotes the communication range of each sensor node)

| GAF-based protocol | Maximum cell size |
|---|---|
| GAF [11] | $\frac{1}{5}R^2$ |
| HGAF [6] | $\frac{1}{2}R^2$ |
| eHGAF [6] | $R^2$ |
| eHGAF with triangle cells [7] | $\frac{3\sqrt{3}}{4}R^2$ |

In [9], we studied a theoretical upper bound on the cell size, and showed that the upper bound is $\pi R^2 - \Delta$, where $R$ denotes the communication range of each sensor node and $\Delta = \{(4\pi - 3\sqrt{3})/6\}R^2$. In [8], we gave an algorithm that uses two types of different cell shapes, and obtained that the cell size can be $\sqrt{3}R^2$ at largest. Since the theoretical upper bound is approximately $1.91R^2$, the cell size obtained in [8] is fairly close to it.

In this paper, we extend the sensor field to 3-dimensional space, and study the power-efficient partitioning of the field into regular polyhedral cells. Further, we prove a theoretical upper bound on the cell size for this problem.

The rest of this paper is organized as follows. Section 2 introduces related works. Section 3 explains the outline of GAF and HGAF. Section 4 extends the sensor field to 3D space, and provides a theoretical upper bound on the cell size. And finally Section 5 offers concluding remarks.

## 2. Related Works

Similarly to GAF, the protocol SPAN saves power consumption by reducing the number of active nodes on the entire network [4]. In SPAN, each node makes its decisions based on the number of neighbor nodes and residual energy, and does not use geographical information. In [12], Liu et al. improved the connectivity among nodes by using honeycomb cells while maintaining the energy efficiency of GAF. In [13], Inagaki et al. extended GAF to HGAF by using a layered structure, and showed extensive simulation results. In [14], Nazrul Alam et al. examined various cell shapes for partitioning 3D sensor field, and studied the network lifetime etc. for each shape comprehensively. In [14], they focused on the number of active nodes as the metrics of energy consumption just as we do here, but their mechanism does not use the hierarchal structure for cell partition like HGAF does.

## 3. Preliminaries

Throughout the paper, as in [6][7][8][9][11], we assume that the radio range of each node is *R* and is unchanged during the operation, and that every node knows its own location information. The detailed explanation about the equations in this section can be found in [11][13].

*3.1 GAF (Geographical Adaptive Fidelity)*

In GAF [11], the entire sensor field is divided into virtual sub-fields called *cells*. In each cell, a node called *active node* is chosen. These active nodes have the following two missions:

(I) The active nodes compose a backbone network for inter-cell data transmissions. Every data across cell-boundaries is conveyed through this backbone in multi-hop manner.

(II) Each active node acts as a gateway node of its own cell. Every transmission across the cell-boundary is via the gateway node.

Each active node is not fixed, and is properly changed over by the other node in the same cell, according to the remaining amount of battery at the time. The election is dynamically performed by the leader-election algorithm [11]. The active nodes are steadily activated, while other nodes are activated only when necessary and are being asleep most of the time. In GAF, nodes are in one of three states: *sleeping*, *discovery*, and *active*. A state transition diagram is shown in Figure 1. Initially, nodes start out in the discovery state. In the discovery state, a node turns on its radio and exchanges discovery messages to find other nodes within the same cell. The discovery message contains a tuple of node id, cell id, estimated node active time, and node state. In the active state, a node plays as an active node for *Ta* time at longest. In the sleeping state, a node turns off its radio and stays in sleep mode for *Ts* time. GAF leaves choices of many parameters including *Td*, *Ta*, node rank, *Ts*, etc. The reader should refer [11] for the details.

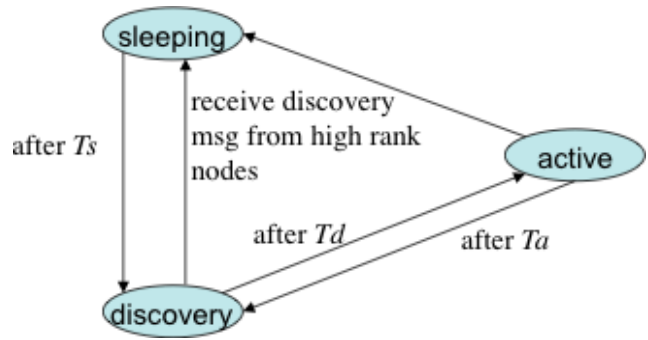

Figure 1. State transitions in GAF

From the viewpoint of energy consumption, it is preferable to make the cell as large as possible [6]. This is because the larger a cell becomes, the smaller the total number of active nodes in the entire sensor field is. The cell size, however, has an upper bound, and we cannot make it larger without limitation. The upper bound is subject to the communication range of each sensor nodes and the following two requirements:

(Req.I) Any pair of active nodes can communicate with each other if their cells are adjacent.

(Req.II) Any active node can communicate with every other node within the cell.

The requirements (Req.I) and (Req.II) are necessary for assuring the missions (I) and (II) of active nodes, respectively.

In GAF, the sensor field is simply divided by square shaped cells of the same size (see Figure 2). Let the size of cell be $r \times r$. For GAF, the requirements (Req.I) and (Req.II) are respectively taken on concrete formulas as follows:

$$r^2 + (2r)^2 \leq R^2, \quad \text{(Req.I-GAF)}$$

$$r^2 + r^2 \leq R^2. \quad \text{(Req.II-GAF)}$$

The inequality (Req.I -GAF) is due to Figure 3(a), and the inequality (Req.II -GAF) is due to Figure 3(b). From these inequalities, we have

$$r \leq \frac{R}{\sqrt{5}},$$

and thus the following claim holds:

**Claim 1.** [11] In GAF, the cell size is bounded above by $(1/5)R^2$.  ∎

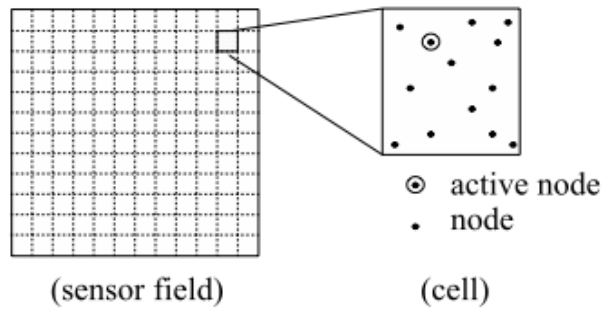

Figure 2. A sensor field divided by square cells

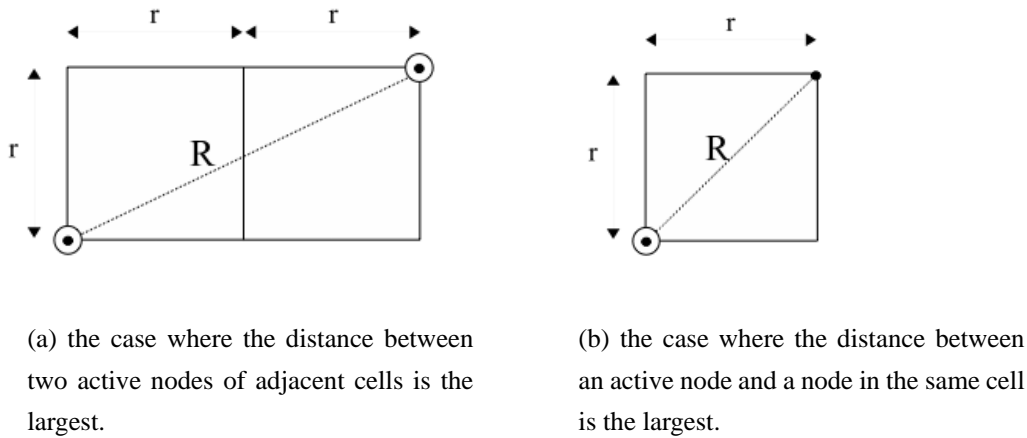

(a) the case where the distance between two active nodes of adjacent cells is the largest.

(b) the case where the distance between an active node and a node in the same cell is the largest.

Figure 3. Examples supporting (Req.I) and (Req.II) for GAF

*3.2 HGAF(Hierarchical Geographical Adaptive Fidelity)*

In HGAF [6], the cell size can be $(1/2)R^2$ at largest, by relaxing the dominant condition (Req.I-GAF) of GAF.

The key idea is to avoid the extreme case illustrated in Figure 3(a). In HGAF, each cell is further divided into smaller squares called *subcells*. A cell of size $r \times r$ is divided into subcells of size $d \times d$. For simplify the exposition, we consider only the case where r is divisible by *d*.

A subcell is called an *active subcell* if it contains an active node of the cell. In HGAF, active subcells are maintained in the same position of the respective cells, and their positions are synchronously rotated (see Figure 4). By this modification, as for (Req.I), we have only to consider the case illustrated in Figure 5. As for (Req.II), the inequality is the same as that of GAF. Hence, we have

$$d^2 + (r+d)^2 \leq R^2, \quad \text{(Req.I-HGAF)}$$

$$r^2 + r^2 \leq R^2. \quad\quad\text{(Req.II-HGAF)}$$

From (Req.I-HGAF), we have

$$r \leq \sqrt{R^2 - d^2} - d,$$

and $r$ can be $R$ at largest when we let $d$ be infinitesimal (i.e., the partition of each cell into subcells is infinitely fine-grained). Hence, the constraint (Req.II-HGAF) becomes the dominant condition here, and thus the following claim holds:

**Claim 2.** [6] In HGAF, the cell size is bounded above by $(1/2)R^2$. ∎

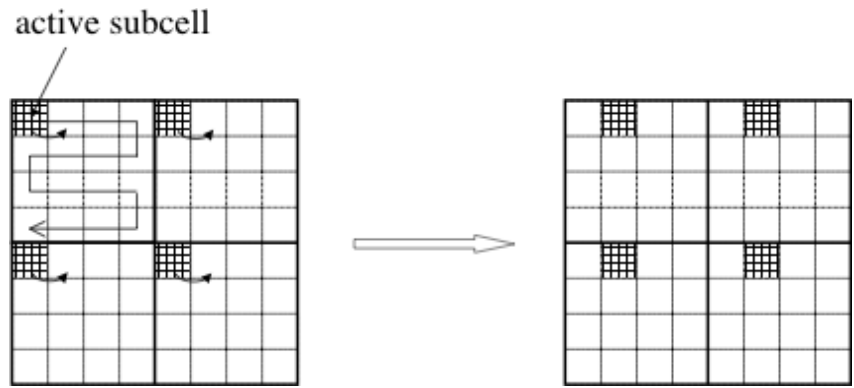

Figure 4. Rotation of active subcells

.

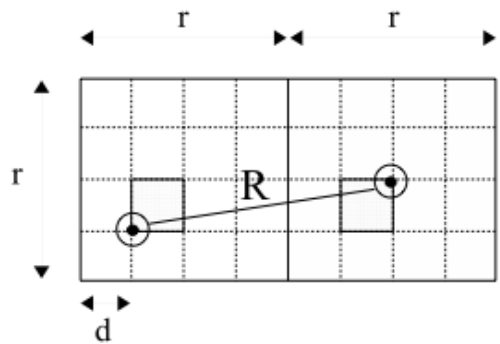

Figure 5. An example supporting (Req.I) for HGAF. Here, the distance between two active nodes of adjacent cells is the largest.

*3.3 eHGAF(extended Hierarchical Geographical Adaptive Fidelity)*

In eHGAF [6], the cell size is improved and can be $R^2$ at largest. Here, the dominant constraint for HGAF, (Req.II-HGAF), is relaxed by keeping active subcells centered. For simplicity, we assume that $r$ is divisible by $d$ and that the quotient is an odd number.

To place an active subcell in the center of its cell, the cell-boundaries are synchronously slid during the operation (see Figure 6). By this modification, for (Req.II), we have only to consider the case illustrated in Figure 7. As for (Req.I), the inequality is the same as that of HGAF. Hence, we have

$$d^2 + (r+d)^2 \leq R^2, \quad \text{(Req.I-eHGAF)}$$

$$2\left(\frac{r+d}{2}\right)^2 \leq R^2. \quad \text{(Req.II-eHGAF)}$$

From (Req.I-eHGAF), $r$ can be $R$ at largest when we let $d$ be infinitesimal. From (Req.II-eHGAF), $r$ can be $\sqrt{2}R$ at largest for infinitesimal $d$. Hence, the constraint (Req.I-eHGAF) becomes the dominant condition here, and thus the following claim holds:

**Claim 3.** [6] In eHGAF, the cell size is bounded above by $R^2$. ∎

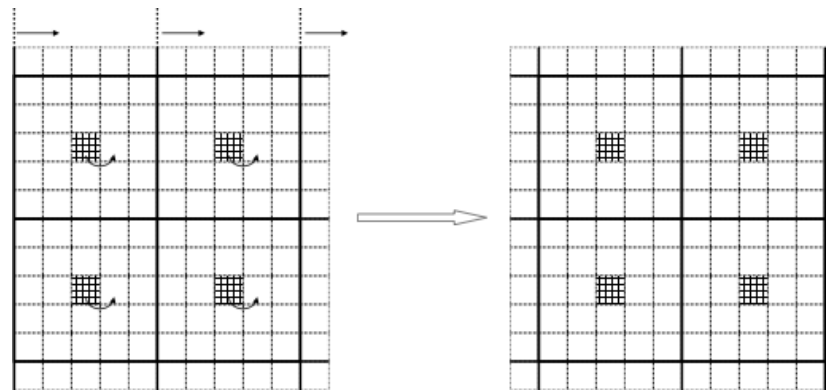

Figure 6. Cell-boundaries are synchronously slid during the operation
so that the active subcells are kept in the center of respective cells.

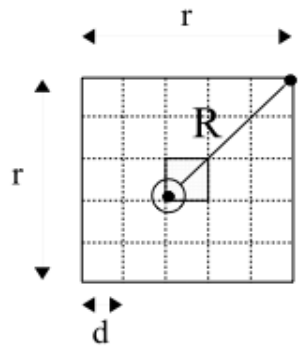

Figure 7. An example supporting (Req.II) for eHGAF. Here, the distance

Network Protocols and Algorithms
ISSN 1943-3581
2009, Vol. 1, No. 2between an active node and a node in the same cell is the largest.

*3.4 eHGAF with Triangle Cells*

In eHGAF with triangle cells [7], the upper bound on the cell size is further improved. The cell size can be approximately $1.299R^2$ at largest, which is 29.9% larger than that of standard eHGAF [6].

The main idea is to change the base-shape of each cell (and subcell) to triangle cells. Previously in GAF, HGAF, and eHGAF, the cells are of square-shape. Although partitioning with squares is regular and natural, a plane can be tiled with other regular polygons such as regular triangle and regular hexagon. In GAF, the cell size can be $\left(1/4\sqrt{3}\right)R^2 \approx 0.144R^2$ if we use triangle cells, and be $\left(3\sqrt{3}/26\right)R^2 \approx 0.200R^2$ if we use hexagon cells. That is, for GAF, we cannot improve the upper bound on the cell size even if we adopt triangle/hexagon cells.

In eHGAF, however, the upper bound on the cell size can be improved up to approximately $1.299\,R^2$ if we use triangle cells. See Figure 8. For simplify the exposition, we assume that *r'* is divisible by *d* and that the quotient is $(3c + 1)$ for some positive integer *c* (this assumption assures that an active subcell can be located in the center (barycenter) of the triangle cell). Here, when *d* is infinitesimal, we can think that the active subcell can be seen as the point just positioned at the barycenter of the regular triangle. In such a case, it is easy to check that the conditions (Req.I) and (Req.II) become identical, and we have the following one inequality:

$$r' \leq \frac{3}{2}R \quad (d \text{ is infinitesimal}).$$

Since *r'* is the height of a regular triangle, the maximum size of triangle-shaped cell is calculated as

$$\frac{1}{2} \cdot \frac{2}{\sqrt{3}} r' \cdot r' = \frac{3\sqrt{3}}{4} R^2 \approx 1.299R^2.$$

Hence, the following claim holds:

**Claim 4.** [7] In eHGAF with triangle cells, the cell size is bounded above by $\left(3\sqrt{3}/4\right)R^2 \approx 1.299R^2$. ∎

92                                                                www.macrothink.org/npa

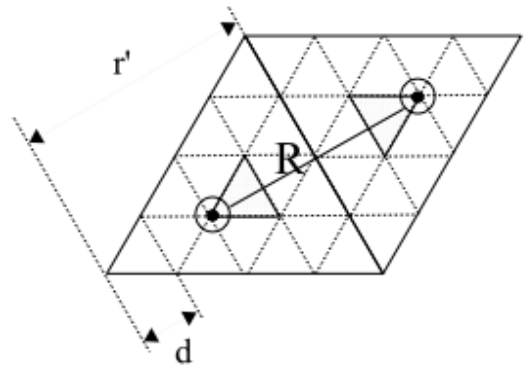

Figure 8. An example supporting (Req.I) for eHGAF with triangle cells. Here, the distance between two active nodes of adjacent cells is the largest.

## 4. Cell Partition of 3D Space

*4.1 Cell Partition of 3D Space by Regular Polyhedra*

In this section, we extend the sensor field to a 3 dimensional space. Here the sensor nodes are distributed over the 3 dimensional sensor field. The 3D sensor network recently attracts attention (e.g. [14][15][16]). Applications of such 3D sensor field include underwater, atmospheric and space applications where height of the network can be significant and nodes are distributed over a 3D space. We examine the energy efficient way of partitioning the field into cells by using regular polyhedra. It is not difficult to confirm that the same strategy used in Section 3 is applicable to the problem here.

In what follows, we consider partitioning the 3D sensor field into cells for eHGAF protocol. Similarly to the argument in Section 3.4, we calculate the upper bound on the cell size with viewing the barycenter of regular polyhedron as the active subcell, assuming that the size of subcells is infinitesimal. To fill up a 3D space with regular polyhedra, we can chose tetra-, hexa-, octa-, dodeca-, and icosahedra.

To begin with, we examine the case of regular hexahedron (regular cube), for it is natural to use cubes for filling up the 3D space. Let $r$ denote the length of an edge of a regular cube (cell). Then, the distance between the barycenter of a cube and that of adjacent cube is given by $r$, and the distance between the barycenter of a cube and a vertex of the cube is $(\sqrt{3}/2)r$ (see Figure 9). Then, for eHGAF with cubic cells, the requirements (Req.I) and (Req.II) are respectively taken on concrete formulas as follows:

$$r \leq R, \qquad \text{(Req.I–hexahedron)}$$

$$\frac{\sqrt{3}}{2} r \leq R. \qquad \text{(Req.II–hexahedron)}$$

Since the volume of the cube is $r^3$, the cell size can be $R^3$ at largest. Hence, the following holds:

**Claim 5**. In eHGAF with hexahedral cells, the cell size is bounded above by $R^3$. ∎

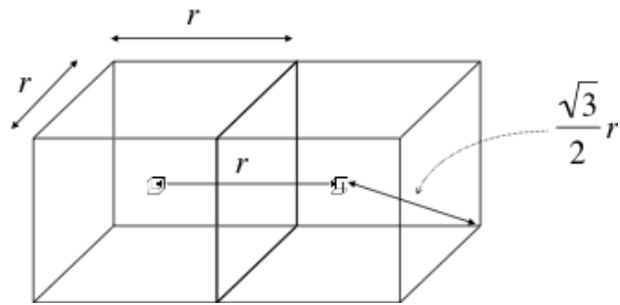

Figure 9. The distance between the barycenter of a cube and that of adjacent cube,

and the distance between the barycenter of a cube and a vertex of the cube.

Next, we examine the case of regular tetrahedron. Let $r$ denote the length of an edge of a regular tetrahedron. Then, the distance between the barycenter of a tetrahedron and that of adjacent one is given by $(1/\sqrt{6})r$, and the distance between the barycenter of a tetrahedron and its vertex is $(\sqrt{6}/4)r$. Then, for eHGAF with tetrahedral cells, the requirements (Req.I) and (Req.II) are respectively taken on concrete formulas as follows:

$$\frac{r}{\sqrt{6}} \leq R, \qquad \text{(Req.I– tetrahedron)}$$

$$\frac{\sqrt{6}}{4}r \leq R. \qquad \text{(Req.II–tetrahedron)}$$

Since the volume of the tetrahedron is $(\sqrt{2}/12)r^3$, the cell size can be $\sqrt{3}R^3$ at largest, and the following claim holds:

**Claim 6**. In eHGAF with regular tetrahedral cells, the cell size is bounded above by $\sqrt{3}R^3 \approx 1.732R^3$. ∎

Similarly, we can calculate upper bounds on the cell size for eHGAF with other polyhedra. The upper bound on the cell size for eHGAF with regular octahedral cells is approximately $0.866R^3$, that of dodecahedral cells is $0.694R^3$, and that of icosahedral cells is $0.627R^3$.

Therefore, we can say that regular tetrahedral cells are preferable for filling up the 3D sensor field.

*4.2 Theoretical Upper Bound on Cell Size*

In this section, we study the upper bound on the cell size for eHGAF with 3D space. We show that the cell size is asymptotically bounded above by $(4\pi/3)R^3 - \Delta$ in average, where $\Delta = (5\pi/12)R^3$. Here, we do not assume the shape of sensor field or that of cells. Please note that $\Delta$ is the volume of the overlapped region of those two spheres of radius $R$ such that the distance between their centers is $R$.

To begin with, we introduce the following two propositions.

**Proposition 1.** If a sensor field consists of a single cell, the size of entire sensor field can be $(4\pi/3)R^3$ at largest. ∎

**Proposition 2.** If a sensor field consists of 2 cells, the size of entire sensor field can be $2(4\pi/3)R^3 - \Delta$ at largest. ∎

An example for Proposition 1 is a sensor field whose shape is a sphere with radius $R$. An example for Proposition 2 is the one whose shape is the union of two spheres such that the radius is $R$ for both spheres and that the distance between their centers is $R$. Please note that $\Delta$ is the volume of the overlapped region of two such spheres.

By generalizing the above two propositions, we can prove the following lemma.

**Lemma 1.** If a sensor field consists of $n$ cells, the size of entire sensor field can be $n(4\pi/3)R^3 - (n-1)\Delta$ at largest. ∎

The Lemma 1 can be proved by mathematical induction on $n$. The base case is due to Proposition 1 and 2. The inductive case can be proved by the following lemma:

**Lemma 2.** Let $S_k$ be a sensor field composed of $k$ cells. Construct $S_{k-1}$ from $S_k$ by the following 3 steps:

(Step 1) choose any single cell $C$ of $S_k$,

(Step 2) remove the sub region of $C$ if it is covered only by the active node of $C$, and

(Step 3) migrate the leftover region of $C$ to the adjacent cell $C'$ if it can be covered by the active node of $C'$.

Then, the following inequality holds:

$$|S_k| - |S_{k-1}| \leq (4\pi/3)R^3 - \Delta,$$

where $|S|$ denotes the volume of $S$. ∎

Lemma 2 can be easily checked by the following observation. Let $C$ denote the cell that is chosen at (Step 1) in Lemma 2. Then, obviously, $|C| \leq (4\pi/3)R^3$ holds. On the other hand, since $C$ has to communicate with at least one cell $C'$ among its adjacent cells, some part of $C$ is covered by the active node of $C'$. So, it is not difficult to confirm that one cannot remove the region whose volume is more than $(4\pi/3)R^3 - \Delta$ at (Step 2) in Lemma 2, and thus Lemma 2 holds.

To prove Lemma 1, let $P(k)$ denote the following proposition:

If a sensor field consists of $k$ cells, the size of entire sensor field can be $k(4\pi/3)R^3 - (k-1)\Delta$ at largest.

As for the proof of inductive case for Lemma 1, if we assume that $P(k)$ holds and that $P(k + 1)$ does not, then we can derive a contradiction by Lemma 2.

95                                                                    www.macrothink.org/npa

From Lemma 1, the average cell size can be calculated as

$$\{n(4\pi/3)R^3 - (n-1)\Delta\}/n.$$

Hence, we can derive the following theorem.

**Theorem 1**. In eHGAF for 3D space, the average cell size can be asymptotically $(4\pi/3)R^3 - \Delta$ at largest. ∎

## 5. Concluding Remarks

In this paper, we showed the following:

1. The cell size of eHGAF for 3D sensor field can be $\sqrt{3}R^3 \approx 1.732R^3$ at largest if we use regular tetrahedral cells.

2. The theoretical upper bound on the cell size of eHGAF for 3D space is asymptotically $(4\pi/3)R^3 - \Delta \approx 2.880R^3$ where $\Delta = (5\pi/12)R^3$.

Here, $R$ is the communication range of each sensor node. Although $1.732R^3$ is far less than $2.880R^3$, it is better to use tetrahedral cells because the cell size would be much smaller if we use other polyhedral cells (e.g., $R^3$ if we use hexahedral cells). Since the total number of active nodes in the entire sensor field inversely relate to the cell size, for eHGAF, we can say that the regular tetrahedral shape is the best for the cell shape if we partition the 3D sensor field with regular polyhedra.

Intuitively, there is a trade-off relation between the cell size and the degree of an active node. For example, if we use tetrahedral cells for partitioning a 3D space, each active node can communicate directly with those active nodes of 4 neighboring cells, which means that the communication range of an active node is overlapped with those of 4 neighboring active nodes. If we use hexahedral (cubic) cells, then the communication range of each active node is overlapped with those of 6 neighboring active nodes, and hence the cell size should be smaller because the portion of redundantly covered area becomes larger. Therefore, if we use a cell shape whereby the degree of an active node is smaller than 4, the cell size can be larger, though the connectivity metrics of the entire network gets worse.

Table 2 summaries the estimated network lifetime for each cell shape. Here, as in [14], we assume that the network lifetime is proportional to the inverse of the number of active nodes in the entire network.

Table 2. Network lifetime for each cell shape.

| Cell shape | Network lifetime compared to the theoretical upper bound |
|---|---|
| tetrahedron | 60.1% |
| hexahedron | 34.7% |
| octahedron | 30.1% |
| dodecahedron | 24.1% |
| icosahedron | 21.8% |
| theoretical upper bound | 100% |

**Acknowledgement**

This work was partly supported by the MEXT Grant-in-Aid for Young Scientists (B) (20700014).

**References**

[1] Chang J.-H., and Tassiulas L. (2000). Energy conserving routing in wireless ad hoc networks, INFOCOM (1), pp.22-31

[2] Kim D., Garcia-Luna-Aceves J., Obraczka K., Cano J., and Manzoni P. (2002). Power-aware routing based on the energy drain rate for mobile ad hoc networks, IEEE International Conference on Computer Communication and Networks

[3] Singh S., Woo M., and Raghavendra C.S. (1998). Power-aware routing in mobile ad hoc networks, Mobile Computing and Networking, pp.181-190

[4] Chen B., Jamieson K., Balakrishnan H., and Morris R. (2001). Span: An energy efficient coordination algorithm for topology maintenance in ad hoc wireless networks, Mobile Computing and Networking, pp.85-96

[5] Heinzelman W., Chandrakasan A., and Balakrishnan H. (2002). An application-specific protocol architecture for wireless microsensor networks, IEEE Transactions on Wireless Communications, 1(4)

[6] Inagaki T., and Ishihara S. (2007). A proposal of a hierarchical power saving technique using location information for sensor networks, IPSJ SIG Technical Report (in Japanese), (14), pp.1-8

[7] Matsumae S., and Miyazaki N. (2008). Improving hierarchical power saving technique using location information for sensor networks, International Conference on Parallel and Distributed Processing Techniques and Applications (PDPTA2008)

[8] Matsumae S. (2008). Hierarchical low power consumption technique with location

information for sensor networks, IASTED International Conference on Communication Systems and Networks (CSN2008)

[9] Matsumae S., and Ooshita F. (2009). Upper bound on cell size for hierarchical GAF, International Conference on Applications and Principles of Information Science (APIS2009)

[10] Xu Y., Heidemann J., and Estrin D. (2000). Adaptive energy conserving routing for multi hop ad hoc networks, Research Report 527, USC/Information Sciences Institute

[11] Xu Y., Heidemann J. S., and Estrin D. (2001). Geography-informed energy conservation for ad hoc routing, Mobile Computing and Networking, pp.70-84

[12] Liu R., Rogers G., and Zhou S. (2006). Honeycomb architecture for energy conservation in wireless sensor networks, GLOBECOM'06

[13] Inagaki T., and Ishihara S. (2009). HGAF: A power saving scheme for wireless sensor networks. IPSJ Journal, 50(10), pp.2520-2531

[14] Nazrul Alam S. M., and Haas Zygmunt J. (2006). Topology control and network lifetime in three-dimensional wireless sensor networks, CoRR abs/cs/0609047

[15] Lee S.-J., Namgung J.-I., and Park S.-H. (2009). Efficient UDD architecture for underwater wireless acoustic sensor network, International Conference on Computational Science and Engineering, pp.972-976

[16] Wang Y., Cao L., Dahlberg T. A., Li F., and Shi X. (2009). Self-organizing fault-tolerant topology control in large-scale three-dimensional wireless networks, ACM Transactions on Autonomous and Adaptive Systems, 4(3)